\documentclass{desyproc}

\usepackage{ifthen} 
\newboolean{uprightparticles}
\setboolean{uprightparticles}{true} 
\usepackage{xspace}
\usepackage{amsfonts}
\usepackage{upgreek}

\newboolean{pdflatex}
\setboolean{pdflatex}{false} 

\def\lhcb {LHCb\xspace}
\def\ux85 {UX85\xspace}
\def\cern {CERN\xspace}
\def\lhc {LHC\xspace}

\ifthenelse{\boolean{uprightparticles}}%
{

 \def\Pmu         {\ensuremath{\upmu}\xspace}

 \def\Ppi         {\ensuremath{\uppi}\xspace}

 \def\Pphi        {\ensuremath{\upphi}\xspace}

 \def\PDelta      {\ensuremath{\Delta}\xspace}                 
 \def\PXi      {\ensuremath{\Xi}\xspace}                 
 \def\PLambda      {\ensuremath{\Lambda}\xspace}                 
 \def\PSigma      {\ensuremath{\Sigma}\xspace}                 
 \def\POmega      {\ensuremath{\Omega}\xspace}                 
 \def\PUpsilon      {\ensuremath{\Upsilon}\xspace}                 
 
 \def\PB      {\ensuremath{\mathrm{B}}\xspace}                 
                  
 \def\PD      {\ensuremath{\mathrm{D}}\xspace}

 \def\PK      {\ensuremath{\mathrm{K}}\xspace}

 \def\Ph      {\ensuremath{\mathrm{h}}\xspace}                 
 \def\Pi      {\ensuremath{\mathrm{i}}\xspace}

 \def\Pp      {\ensuremath{\mathrm{p}}\xspace}

}
{

 \def\Pmu         {\ensuremath{\mu}\xspace}

 \def\Ppi         {\ensuremath{\pi}\xspace}

 \def\Pphi        {\ensuremath{\phi}\xspace}

 \mathchardef\PDelta="7101
 \mathchardef\PXi="7104
 \mathchardef\PLambda="7103
 \mathchardef\PSigma="7106
 \mathchardef\POmega="710A
 \mathchardef\PUpsilon="7107
                  
 \def\PB      {\ensuremath{B}\xspace}                 
                  
 \def\PD      {\ensuremath{D}\xspace}

 \def\PK      {\ensuremath{K}\xspace}

 \def\Ph      {\ensuremath{h}\xspace}                 
 \def\Pi      {\ensuremath{i}\xspace}

 \def\Pp      {\ensuremath{p}\xspace}

}

\def\mmu        {\ensuremath{\Pmu}\xspace}

\def\mun        {\ensuremath{\Pmu^-}\xspace} 

\def\c     {\ensuremath{\Pc}\xspace}

\def\pion  {\ensuremath{\Ppi}\xspace}

\def\pip   {\ensuremath{\pion^+}\xspace}
\def\pim   {\ensuremath{\pion^-}\xspace}
\def\pipi  {\ensuremath{\pion^+\pion^-}\xspace}
\def\pipm  {\ensuremath{\pion^\pm}\xspace}

\def\kaon  {\ensuremath{\PK}\xspace}
  \def\Kbar  {\kern 0.2em\overline{\kern -0.2em \PK}{}\xspace}

\def\Kz    {\ensuremath{\kaon^0}\xspace}
\def\Kzb   {\ensuremath{\Kbar^0}\xspace}
\def\KzKzb {\ensuremath{\Kz \kern -0.16em \Kzb}\xspace}
\def\Kp    {\ensuremath{\kaon^+}\xspace}
\def\Km    {\ensuremath{\kaon^-}\xspace}
\def\Kpm   {\ensuremath{\kaon^\pm}\xspace}
\def\Kmp   {\ensuremath{\kaon^\mp}\xspace}
\def\KpKm  {\ensuremath{\Kp \kern -0.16em \Km}\xspace}
\def\KS    {\ensuremath{\kaon^0_{\rm\scriptscriptstyle S}}\xspace}

  \def\Dbar    {\kern 0.2em\overline{\kern -0.2em \PD}{}\xspace}
\def\D       {\ensuremath{\PD}\xspace}

\def\Dz      {\ensuremath{\D^0}\xspace}
\def\Dzb     {\ensuremath{\Dbar^0}\xspace}
\def\DzDzb   {\ensuremath{\Dz {\kern -0.16em \Dzb}}\xspace}
\def\Dp      {\ensuremath{\D^+}\xspace}
\def\Dm      {\ensuremath{\D^-}\xspace}
\def\Dpm     {\ensuremath{\D^\pm}\xspace}

\def\DpDm    {\ensuremath{\Dp {\kern -0.16em \Dm}}\xspace}

\def\Dstarp  {\ensuremath{\D^{*+}}\xspace}

\def\Dspm    {\ensuremath{\D^{\pm}_s}\xspace}

  \def\Bbar    {\kern 0.18em\overline{\kern -0.18em \PB}{}\xspace}

  \def\Y#1S{\ensuremath{\PUpsilon{(#1S)}}\xspace}

\def\proton      {\ensuremath{\Pp}\xspace}

\newcommand{\decay}[2]{\ensuremath{#1\!\to #2}\xspace}         

\def\to                 {\ensuremath{\rightarrow}\xspace}

\def\order   {\ensuremath{\mathcal{O}}\xspace}

\def\CP                {\ensuremath{C\!P}\xspace}

\def\AT#1     {\ensuremath{A_T^{#1}}\xspace}           

\def\C#1      {\ensuremath{\mathcal{C}_{#1}}\xspace}                       
\def\Cp#1     {\ensuremath{\mathcal{C}_{#1}^{'}}\xspace}                    
\def\Ceff#1   {\ensuremath{\mathcal{C}_{#1}^{\mathrm{(eff)}}}\xspace}        
\def\Cpeff#1  {\ensuremath{\mathcal{C}_{#1}^{'\mathrm{(eff)}}}\xspace}       
\def\Ope#1    {\ensuremath{\mathcal{O}_{#1}}\xspace}                       
\def\Opep#1   {\ensuremath{\mathcal{O}_{#1}^{'}}\xspace}                    

\def\agamma     {\ensuremath{A_{\Gamma}}\xspace}

\def\kk         {\ensuremath{\PK\PK}\xspace}

\newcommand{\ket}[1]{\ensuremath{|#1\rangle}}              

\newcommand{\tev}{\ensuremath{\mathrm{\,Te\kern -0.1em V}}\xspace}
\newcommand{\gev}{\ensuremath{\mathrm{\,Ge\kern -0.1em V}}\xspace}
\newcommand{\mev}{\ensuremath{\mathrm{\,Me\kern -0.1em V}}\xspace}
\newcommand{\kev}{\ensuremath{\mathrm{\,ke\kern -0.1em V}}\xspace}
\newcommand{\ev}{\ensuremath{\mathrm{\,e\kern -0.1em V}}\xspace}
\newcommand{\gevc}{\ensuremath{{\mathrm{\,Ge\kern -0.1em V\!/}c}}\xspace}
\newcommand{\mevc}{\ensuremath{{\mathrm{\,Me\kern -0.1em V\!/}c}}\xspace}
\newcommand{\gevcc}{\ensuremath{{\mathrm{\,Ge\kern -0.1em V\!/}c^2}}\xspace}
\newcommand{\gevgevcccc}{\ensuremath{{\mathrm{\,Ge\kern -0.1em V^2\!/}c^4}}\xspace}
\newcommand{\mevcc}{\ensuremath{{\mathrm{\,Me\kern -0.1em V\!/}c^2}}\xspace}

\def\invfb   {\ensuremath{\mbox{\,fb}^{-1}}\xspace}

\def\order{{\ensuremath{\cal O}}\xspace}
\newcommand{\chisq}{\ensuremath{\chi^2}\xspace}

\def\gsim{{~\raise.15em\hbox{$>$}\kern-.85em
          \lower.35em\hbox{$\sim$}~}\xspace}
\def\lsim{{~\raise.15em\hbox{$<$}\kern-.85em
          \lower.35em\hbox{$\sim$}~}\xspace}

\def\sqs   {\ensuremath{\protect\sqrt{s}}\xspace}

\def\ptot       {\mbox{$p$}\xspace}

\def\tell1  {TELL1\xspace}
\def\ukl1   {UKL1\xspace}

\def\hphm       {\ensuremath{\Ph^+ \Ph^-}\xspace}

\def\gammahat   {\ensuremath{\hat{\Gamma}}\xspace}

\def\agammadefnot  {\ensuremath{\frac{\gammahat(\Dz \to f) - \gammahat(\Dzb \to f)}{\gammahat(\Dz \to f) + \gammahat(\Dzb \to f)}}\xspace}
\def\etacp      {\ensuremath{\eta_{\CP}}\space}
\def\agammaexp  {\ensuremath{\etacp\left[\frac{1}{2}(A_m+A_d)y\cos\phi-x\sin\phi\right]}\xspace}

\def\dzkk       {\decay{\Dz}{\Kp\Km}}
\def\dzpipi     {\decay{\Dz}{\pip\pim}}

\def\dzkpi      {\decay{\Dz}{\Km\pip}}

\def\dstdzpi    {\decay{\Dstarp}{\Dz\pip_s}\xspace}
\def\dkpidcs    {\decay{\Dz}{\Kp\pim}}

\def\SlowPi  		{\ensuremath{\pion_s}\xspace}

\def\c          {\ensuremath{c}\xspace}

\newcommand{\magnitude}[1]{\ensuremath{\left|{#1}\right|}\xspace}
\newcommand{\magsq}[1]{\ensuremath{\magnitude{#1}^2}\xspace}

\newcommand{\orderof}[1]{\ensuremath{\order(#1)}\xspace}

\newcommand{\tene}[1]{\ensuremath{10^{#1}}\xspace}
\newcommand{\xtene}[2]{\ensuremath{#1 \times \tene{#2}}\xspace}
\newcommand{\otene}[1]{\orderof{\tene{#1}}}

\def\onehalf		{\ensuremath{\frac{1}{2}}\xspace}

\def\CPV{\CP-violation\xspace}
\def\C	{\ensuremath{C}\xspace}

\def\T	{\ensuremath{T}\xspace}

\def\CPV     {\CP violation\xspace}

\def\pp      {\ensuremath{\proton\proton}\xspace}

\def\dzkkpipi{\decay{\Dz}{\KpKm\pipi}}
\def\dpmopts{\ensuremath{\Dpm_{(s)}}\xspace}
\def\doptsksh{\decay{\dpmopts}{\KS h^{\pm}}}
\def\doptspksh{\decay{\Dp_{(s)}}{\KS h^{+}}}
\def\doptsmksh{\decay{\Dm_{(s)}}{\KS h^{-}}}
\newcommand{\Asym}[2]{\ensuremath{{\cal A}^{#1}_{#2}}\xspace}
\newcommand{\Ameas}[1]{{\Asym{#1}{\text{meas}}}}
\newcommand{\Acp}[1]{\Asym{#1}{CP}}
\newcommand{\Adet}[1]{{\Asym{#1}{\text{det}}}}
\newcommand{\Aprod}[1]{{\Asym{#1}{\text{prod}}}}

\newcommand{\cpasymflat}[2]{\ensuremath{(#1 - #2)/(#1 + #2)}\xspace}
\def\ACPdoptsksh{\Acp{\doptsksh}}
\def\Ameasdoptsksh{\Ameas{\doptsksh}}
\def\dpmksk{\decay{\Dpm}{\KS\Kpm}}
\def\dpmkspi{\decay{\Dpm}{\KS\pipm}}
\def\dspmkspi{\decay{\Dspm}{\KS\pipm}}
\def\dspmksk{\decay{\Dspm}{\KS\Kpm}}
\def\dspmphipi{\decay{\Dspm}{\Pphi\pipm}}
\def\ADD{\Asym{DD}{CP}}
\def\AKS{\Asym{}{\KS}}
\def\Ameasdskspi{\Ameas{\dspmkspi}}
\def\Ameasdsksk{\Ameas{\dspmksk}}
\def\Ameasdkspi{\Ameas{\dpmkspi}}
\def\Ameasdksk{\Ameas{\dpmksk}}
\def\Acpdskspi{\Acp{\dspmkspi}}
\def\Acpdksk{\Acp{\dpmksk}}
\def\Ameasdsphipi{\Ameas{\dspmphipi}}

\def\ct{\ensuremath{C_T}\xspace}
\def\ctb{\ensuremath{\bar{C}_T}\space}
\newcommand{\vecp}[1]{\ensuremath{\vec{\ptot}_{#1}}\xspace}
\def\At{\ensuremath{A_T}\xspace}
\def\Atb{\ensuremath{\bar{A}_T}\xspace}
\def\atodd{\ensuremath{a^{\T\text{-odd}}_{\CP}}\xspace}
\def\btodzmux{\decay{\PB}{\Dz \mun X}}
\def\dppipipi{\decay{\Dp}{\pim\pip\pip}}
\def\dzpipipipi{\decay{\Dz}{\pipi\pipi}}

\def\dzhphm{\decay{\Dz}{\hphm}}
\def\Ameasdzhh{\Ameas{\dzhphm}}
\def\Acpdzhh{\Acp{\dzhphm}}
\def\Ameasdzpipi{\Ameas{\dzpipi}}
\def\Ameasdzkk{\Ameas{\dzkk}}
\def\Acpdzpipi{\Acp{\dzpipi}}
\def\Acpdzkk{\Acp{\dzkk}}
\def\DACP{\ensuremath{\Delta\Acp{}{}}\xspace}
\def\Ameasdzkpi{\Ameas{\dzkpi}}

\begin{document}
\title{Measurements of \CP Violation and Mixing in Charm Decays at \lhcb}

\author{{\slshape Michael Alexander}, on behalf of the \lhcb collaboration\\[1ex]
University of Glasgow, University Avenue, Glasgow, UK, G12 8QQ}

\contribID{244}

\confID{8648}  
\desyproc{DESY-PROC-2014-04}
\acronym{PANIC14} 
\doi  

\maketitle

\begin{abstract}
  During run I, the \lhcb experiment at the \lhc, \cern, collected 1.0 \invfb of \pp collisions 
  at $\sqs = 7 \tev$ and 2.0 \invfb at $\sqs = 8 \tev$, yielding
  the world's largest sample of decays of charmed hadrons. This sample is used to 
  search for direct and indirect \CPV in charm and to measure \Dz mixing parameters. 
  Recent measurements from several complementary decay modes are presented.
\end{abstract}

\section{Introduction}
\label{sec:intro}
The \lhcb detector is a forward-arm spectrometer, with pseudo-rapidity coverage $2 < \eta < 5$, specifically designed for high precision measurements of decays of $b$ and \c hadrons \cite{JINST_LHCb}. During run I, the experiment collected 1.0 \invfb of \pp collisions at $\sqs = 7 \tev$ and 2.0 \invfb at $\sqs = 8 \tev$, yielding the world's largest sample of decays of charmed hadrons. This allows \CPV and mixing in charm to be studied with unprecedented precision in many complementary decay modes. The Standard Model (SM) predicts \CP asymmetries to be \otene{-3} or less in charm interactions \cite{Brod_directCPVInSCSCharm2012, Bobrowski_indirectCPVCharm2010}; observation of significantly larger \CP violating effects could indicate new physics.

\begin{sloppypar}
For a decay \decay{\PD}{f} and its \CP conjugate \decay{\bar{\PD}}{\bar{f}}, with amplitudes $A_f$ and $\bar{A}_{\bar{f}}$ respectively, direct \CP violation is quantified by
\mbox{$A_d = \cpasymflat{\magsq{A_f}}{\magsq{\bar{A}_{\bar{f}}}}$}.
For \Dz mesons, the mass eigenstates \ket{\PD_{1,2}}, with masses $m_{1,2}$ and widths $\Gamma_{1,2}$, are defined in terms of the flavour eigenstates, \ket{\Dz} and \ket{\Dzb}, as
\mbox{$\ket{\PD_{1,2}} = p \ket{\Dz} \pm q \ket{\Dzb}$},
with $p$ and $q$ complex, satisfying \mbox{$\magsq{p} + \magsq{q} = 1$}. The rate of mixing is quantified by 
\mbox{$x \equiv 2(m_2 - m_1)/(\Gamma_1 + \Gamma_2)$} and \mbox{$y \equiv (\Gamma_2 - \Gamma_1)/(\Gamma_1 + \Gamma_2)$}.
\CPV in mixing is quantified by 
\mbox{$A_m\equiv(|q/p|^2-|p/q|^2)/(|q/p|^2+|p/q|^2)$} 
and the interference between mixing and decay (when $f = \bar{f}$) by 
\mbox{$\lambda_f \equiv q\bar{A}_{f}/p A_f = \magnitude{q\bar{A}_{f}/p A_f} e^{i \phi}$}.
\end{sloppypar}

The flavour of the \Dz meson at production is determined using either \dstdzpi decays, where the charge of the ``soft pion'', \SlowPi, track gives the \Dz flavour, or \btodzmux decays, where the charge of the \mmu track gives the \Dz flavour. 

\section{Multi-body \PD decays}
\label{sec:multibody}

Multi-body \PD decays are sensitive to \CPV due to the interference of different resonances across the multi-body phase space. In \dzkkpipi decays, triple products of final state particle momenta in the \Dz rest frame, defined as \mbox{$\ct \equiv \vecp{\Kp} \cdot \left(\vecp{\pip} \times \vecp{\pim}\right)$} and \mbox{$\ctb \equiv \vecp{\Km} \cdot \left(\vecp{\pim} \times \vecp{\pip}\right)$}
for \Dz and \Dzb mesons respectively,
are odd under \T. The decay rate asymmetries 
\begin{align*}
\At & \equiv \cpasymflat{\Gamma(\ct > 0)}{\Gamma(\ct < 0)}, \\
\Atb & \equiv \cpasymflat{\Gamma(-\ctb > 0)}{\Gamma(-\ctb < 0)},
\end{align*}
 are thus sensitive to \CPV. However, final state interactions introduce significant asymmetries, and so the difference $\atodd \equiv \onehalf (\At - \Atb)$ is used to access the asymmetry of the \Dz meson. The observable \atodd is by definition insensitive to production and detection asymmetries, so is very robust against systematic uncertainties.

Using 3 \invfb of data, the phase space integrated measurements are found to be \cite{lhcb_toddfourbody2014}
\begin{align*}
  \At &= (-7.18 \pm 0.41 (\text{stat}) \pm 0.13 (\text{syst})) \%, \\
  \Atb &= (-7.55 \pm 0.41 (\text{stat}) \pm 0.12 (\text{syst})) \%, \\
  \atodd &= (0.18 \pm 0.29 (\text{stat}) \pm 0.04 (\text{syst})) \%.
\end{align*}
This shows no evidence of \CPV but achieves a significant improvement in precision over the previous world average of $\atodd = (0.11 \pm 0.67)$\cite{HFAG2014}.

The same measurements are also performed in 32 bins of Cabibbo-Maksimowicz phase space variables, defined as the invariant mass squared of the \pipi (\KpKm) pair, $m^2_{\pipi}$ ($m^2_{\KpKm}$); the cosine of the angle of the \pip (\Kp) with respect to the direction opposite to the \Dz momentum in the \pipi (\KpKm) rest frame, $\cos(\theta_{\pion})$ ($\cos(\theta_{\kaon})$); and the angle between the \KpKm and \pipi planes in the \Dz rest frame, $\phi$. The asymmetries are extracted in each bin of phase space and \atodd calculated. A \chisq test for consistency across the phase space is performed, yielding a p-value of \mbox{74 \%}. Similarly, binning in the decay time of the \Dz candidates and performing the same test gives sensitivity to indirect \CPV. This yields a p-value of \mbox{72 \%}, so there is no evidence for direct or indirect \CPV. 

\begin{sloppypar}
A complementary method for studying \CPV in multi-body \PD meson decays is to examine \CP asymmetries across the multi-body phase space directly. Signal yields are obtained in bins of the multi-body phase space, and the test statistic 
\mbox{$S^i_{CP} \equiv (N_i(\Dz) - \alpha N_i(\Dzb))/\sqrt{\alpha (N_i(\Dz) + N_i(\Dzb))}$}, calculated in each bin $i$, where \mbox{$\alpha \equiv N(\Dz)/N(\Dzb)$} cancels any global production and detection asymmetries. A \chisq test for consistency with zero \CPV is performed using $\chisq = \Sigma_i {S^i_{CP}}^2$ and $N_{bins} - 1$ degrees of freedom. This analysis has been performed on \dzkkpipi and \dzpipipipi candidates, using 1 \invfb of data, for which the nominal binning scheme yields a $p$-value of \mbox{9.1 \%} (\mbox{41 \%}) for \dzkkpipi (\dzpipipipi) \cite{lhcb_mirandafourbody2013}. The decay \dppipipi has also been studied in this way, using 1 \invfb of data \cite{lhcb_mirandathreebody2013}. Various binning schema are used, as well as an unbinned method to measure \CP asymmetries, all of which yield p-values of more than \mbox{20 \%}. Thus, no evidence for \CPV is found in these decay modes.
\end{sloppypar}

\section{\CP violation in \doptsksh}
\label{sec:d2ksh}

The singly-Cabibbo-suppressed (SCS) decays \dpmksk and \dspmkspi offer a means of measuring direct \CPV with high precision. The \CP asymmetry is defined as \\ \mbox{$\ACPdoptsksh \equiv \cpasymflat{\Gamma(\doptspksh)}{\Gamma(\doptsmksh)}$}, while the measured asymmetry is
\begin{align*}
    \Ameasdoptsksh & \equiv \cpasymflat{N^{\doptspksh}_{\text{sig}}}{N^{\doptsmksh}_{\text{sig}}} \\
    & \simeq \ACPdoptsksh + \Aprod{\dpmopts} + \Adet{h^\pm} + \AKS.
\end{align*}
Here $N_{\text{sig}}$ is the number of signal candidates of the given decay, \Aprod{\dpmopts} is the production asymmetry of the \dpmopts meson, \Adet{h^\pm} is the detection asymmetry of the $h^{\pm}$ meson, and \AKS is the combined detection and \CP asymmetry of the \KS meson. Assuming negligible \CP violation in the Cabibbo-favoured (CF) decays \dspmksk, \dpmkspi and \dspmphipi, the production and detection asymmetries cancel in the double difference
\begin{align*}
  \ADD & \equiv \left[\Ameasdskspi - \Ameasdsksk\right] - \left[\Ameasdkspi - \Ameasdksk\right] - 2\AKS, \\
  & = \Acpdksk + \Acpdskspi,
\end{align*}
while the \KS asymmetry is calculable, so the sum of the \CP asymmetries can be measured. Similarly the individual \CP asymmetries can be accessed using
  \begin{align*}
    \Acpdksk & = \left[\Ameasdksk - \Ameasdsksk\right] - \left[\Ameasdkspi - \Ameasdsphipi\right] - \AKS, \\
    \Acpdskspi &= \Ameasdskspi - \Ameasdsphipi - \AKS.
  \end{align*}
Using 3 \invfb of data the results thus obtained are \cite{lhcb_d2ksh2014}
\begin{align*}
  \Acpdksk + \Acpdskspi & = (+0.41 \pm 0.49(\text{stat}) \pm 0.26(\text{syst})) \%, \\
  \Acpdksk & = (+0.03 \pm 0.17(\text{stat}) \pm 0.14(\text{syst})) \%, \\
  \Acpdskspi & = (+0.38\pm0.46(\text{stat}) \pm 0.17(\text{syst})) \%.
\end{align*}
These are the most precise measurements of their kind to date and show no evidence of \CPV.

\section{Mixing and \CPV in \decay{\Dz}{\Ph^+\Ph^{(\prime)-}} decays}
\label{sec:d2hh}

Decays of \decay{\Dz}{\Ph^+\Ph^{(\prime)-}} provide a means of measuring direct and indirect \CPV, as well as mixing, in the \Dz system. The measured \CP asymmetry in \dzkk and \dzpipi decays, flavour tagged using \btodzmux decays, is $\Ameasdzhh = \Acpdzhh + \Adet{\mmu^\pm} + \Aprod{\PB}$. The \pipi and \KpKm final states are \CP eigenstates, so have no detection asymmetry. Defining $\DACP \equiv \Ameasdzkk - \Ameasdzpipi = \Acpdzkk - \Acpdzpipi$, the nuisance asymmetries cancel. Similarly to the analysis described in Sec. \ref{sec:d2ksh}, CF decays can be used to cancel nuisance asymmetries as $\Acpdzkk = \Ameasdzkk - \Ameasdzkpi + \Adet{\Kmp\pipm}$, and \Adet{\Kmp\pipm} can be calculated using the asymmetries of \decay{\Dp}{\Km\pip\pip} and \decay{\Dp}{\KS\pip} decays, and the known \AKS. The asymmetry  \Acpdzpipi can then be determined using $\Acpdzpipi = \Acpdzkk - \DACP$.

Using 3 \invfb of data yields \cite{lhcb_sldacp2014}
\begin{align*}
  \DACP & = (+0.14 \pm 0.16 (\text{stat}) \pm 0.08 (\text{syst})) \%, \\
  \Acpdzkk & = (-0.06 \pm 0.15 (\text{stat}) \pm 0.10 (\text{syst})) \%,\\
  \Acpdzpipi & = (-0.20 \pm 0.19 (\text{stat}) \pm 0.10 (\text{syst})) \%.
\end{align*}

Indirect \CPV in \dzkk and \dzpipi decays can be measured using 
  \begin{align*}
    \agamma \equiv \agammadefnot \approx \agammaexp.
  \end{align*}
Here, \gammahat is the inverse of the effective lifetime of the decay and \etacp is the \CP eigenvalue of $f$. The effective lifetimes are measured directly using a data-driven, per-candidate correction for the selection efficiency on 1 \invfb of data, yielding \cite{lhcb_agamma2014}
\begin{align*}
  \agamma(\pion\pion) &= (+0.033 \pm 0.106(\text{stat}) \pm 0.014(\text{syst})) \%, \\
  \agamma(\kk) &= (-0.035 \pm 0.062(\text{stat}) \pm 0.012(\text{syst})) \%.
\end{align*}
Thus, no evidence for direct or indirect \CPV in \decay{\Dz}{\hphm} decays is found.

Mixing in the \Dz system is measured using the ratio of the decay rates of ``wrong sign'' DCS \dkpidcs to ``right sign'' CF \dzkpi as a function of \Dz decay time, as
\begin{equation*}
  R(t) = \frac{N_{WS}(t)}{N_{RS}(t)} = R_D + \sqrt{R_D} y^{\prime} t + \frac{x^{\prime 2} + y^{\prime 2}}{4} t^2,
\end{equation*}
where $R_D = \magsq{\frac{A_{DCS}}{A_{CF}}}$, $x^{\prime} = x \cos(\delta) + y \sin(\delta)$, $y^{\prime} = - x \sin(\delta) + y \cos(\delta)$, and $\delta = \mathrm{arg}\left(\frac{A_{DCS}}{A_{CF}}\right)$. Using 3 \invfb of data yields \cite{lhcb_charmmixing2013}
\begin{equation*}
  x^{\prime 2} = \xtene{(5.5\pm4.9)}{-5}, y^{\prime} = \xtene{(4.8\pm1.0)}{-3}, R_D = \xtene{(3.568\pm0.066)}{-3}.
\end{equation*}
Allowing for \CP violation yields:
\begin{equation*}
  A_D \equiv (R_D(\Dz) - R_D(\Dzb))/(R_D(\Dz) - R_D(\Dzb)) = (-0.7 \pm 1.9) \%,
\end{equation*}
\begin{equation*}
  0.75 < \magnitude{q/p} < 1.24, (\text{68.3 \% CL}).
\end{equation*}
These are the most precise measurements of mixing in the \Dz system and of \CP violation in \dkpidcs decays to date.

\section{Conclusions}

There is a rich programme of charm physics studies at the \lhcb experiment, with many complementary measurements already performed using some or all the 3 \invfb of data collected during run I. No evidence for \CPV has been found, though constraints of \otene{-3} have been achieved in many decay modes. Mixing in the \Dz system has also been measured to unprecedented precision. With run II shortly to begin, and the \lhcb upgrade in the near future, there are great prospects for future measurements with precisions of \otene{-4}, which will tightly constrain, or potentially discover, new physics.
 

\begin{footnotesize}

\bibliographystyle{unsrt}
\bibliography{bibliography}

\end{footnotesize}


\end{document}